\documentclass[pra,twocolumn,showpacs]{revtex4-1}

\usepackage{epsfig}
\usepackage{bm}
\usepackage[matrix,frame,arrow]{xy}

\usepackage[applemac]{inputenc}
\usepackage[T1]{fontenc}
\usepackage{lmodern}
\usepackage[english]{babel}
\usepackage{amsmath}
\usepackage{ae}
\usepackage{units}
\usepackage{amssymb}
\usepackage{icomma}
\usepackage{color}
\usepackage{graphicx}
\usepackage{bbm}
\usepackage{url}
\usepackage{nomencl}
\usepackage{subfigure}
\usepackage{tikz}
\usepackage{slashed}

\usepackage{fixmath}
\usepackage{booktabs}

\newcommand{\rd}{\ensuremath{\mathrm{d}}}
\newcommand{\id}{\ensuremath{\,\rd}}

\newcommand{\ket}[1]{|#1\rangle}  

\newcommand{\ketbra}[2]{\left| #1 \rangle \langle #2 \right|}
\newcommand{\brakket}[3]{\left\langle #1\left| #2 \right| #3\right\rangle}
\newcommand{\expec}[1]{\left\langle #1 \right\rangle}

\newcommand{\comm}[2]{\left[ #1, #2 \right]}
\newcommand{\lind}[1]{\mathcal{D}\left[#1\right]}

\newcommand{\sz}{\sigma_z}
\newcommand{\sm}{\sigma_-}
\renewcommand{\sp}{\sigma_+}
\newcommand{\smm}{\sigma_-^m}
\newcommand{\spm}{\sigma_+^m}

\newcommand{\abssq}[1]{\left| #1 \right|^2}

\newcommand{\im}{\text{Im}}

\newcommand{\nn}{\nonumber}

\newcommand{\figref}[1]{\mbox{Fig.~\ref{#1}}}

\newcommand{\secref}[1]{\mbox{Sec.~\ref{#1}}}

\newcommand{\appref}[1]{\mbox{Appendix~\ref{#1}}}
\renewcommand{\eqref}[1]{\mbox{Eq.~(\ref{#1})}}

\newcommand{\be}{\begin{equation}}
\newcommand{\ee}{\end{equation}}
\newcommand{\bea}{\begin{eqnarray}}
\newcommand{\eea}{\end{eqnarray}}

\begin{document}

\title{Designing frequency-dependent relaxation rates and Lamb shift \\ for a giant artificial atom}

\author{Anton \surname{Frisk Kockum}}
\email[e-mail:]{friska@chalmers.se}
\author{Per Delsing}
\author{G\"{o}ran Johansson}
\email[e-mail:]{goran.l.johansson@chalmers.se}

\affiliation{Department of Microtechnology and Nanoscience$,$ MC2$,$\\ Chalmers\:University\:of\:Technology$,$ SE-412 96\:Gothenburg$,$ Sweden}

\date{\today}

\begin{abstract}
In traditional quantum optics, where the interaction between atoms and light at optical frequencies is studied, the atoms can be approximated as point-like when compared to the wavelength of light. So far, this relation has also been true for artificial atoms made out of superconducting circuits or quantum dots, interacting with microwave radiation. However, recent and ongoing experiments using surface acoustic waves show that a single artificial atom can be coupled to a bosonic field at several points wavelengths apart. Here, we theoretically study this type of system. We find that the multiple coupling points give rise to a frequency dependence in the coupling strength between the atom and its environment, and also in the Lamb shift of the atom. The frequency dependence is given by the discrete Fourier transform of the coupling point coordinates and can therefore be designed. We discuss a number of possible applications for this phenomenon, including tunable coupling, single-atom lasing, and other effects that can be achieved by designing the relative coupling strengths of different transitions in a multi-level atom.

\end{abstract}

\pacs{03.65.Yz, 42.50.-p, 77.65.Dq, 84.40.Az}

\maketitle


\section{Introduction}

Atoms found in nature are so small ($r \approx \unit[10^{-10}]{m}$) that they in most cases can be approximated as point-like. This is certainly the case in traditional quantum optics, which is concerned with the interaction between such atoms and electromagnetic light at optical wavelengths ($\lambda \approx \unit[10^{-6} - 10^{-7}]{m}$) \cite{LeibfriedRMP2003,MillerJPB2005}. Atoms excited to high Rydberg states can reach comparable sizes ($r \approx \unit[10^{-8} - 10^{-7}]{m}$), but in experiments they interact with microwave radiation ($\lambda \approx \unit[10^{-3} - 10^{-1}]{m}$) \cite{HarocheRMP2013,WaltherRPP2006}. 

In recent years, many research groups have started building ``artificial atoms'' using, \emph{e.g.}, superconducting circuits \cite{ClarkeNature2008} or quantum dots \cite{HansonRMP2007}. These artificial atoms can be designed to have various desirable properties such as (tunable) strong coupling strengths \cite{WallraffNature2004,SrinivasanPRL2011} and specific (tunable) level structures, which can be an advantage compared to working with real atoms with fixed properties set by nature. Since the artificial atoms can be made to interact with microwave radiation \cite{WallraffNature2004,BlaisPRA2004,FreyPRL2012}, they realize "quantum optics on a chip", also referred to as circuit quantum electrodynamics (cQED). The advantages of cQED has been demonstrated by tests of quantum optics theories in new regimes not previously accessible \cite{WilsonNature2011,AstafievScience2010}.

Even though the circuits making up the artificial atoms can be quite large ($l \approx \unit[10^{-4}]{m}$), they are still effectively point-like when compared to the wavelength of the microwave fields they interact with. However, a few very recent experiments show that this need not always be the case. For example, there is ongoing work \cite{GustafssonApr2014,GustafssonNatPhys2012} on coupling a certain type of artificial atom, a superconducting circuit called transmon \cite{KochPRA2007}, to surface acoustic waves (SAWs) \cite{DattaSAWBook,MorganSAWBook}. Due to the low SAW velocity, the transmon is several phonon wavelengths ($\lambda \approx \unit[10^{-6}]{m}$) long in this experiment, making it a giant artificial atom. Also, a recent update of the transmon design \cite{BarendsPRL2013} opens up the possibility of coupling it at several points, wavelengths apart, to a meandering microwave transmission line. Furthermore, 3D transmons are approaching wavelength sizes \cite{KirchmairNature2013}.

\begin{figure*}[ht!]
\includegraphics[width=2\columnwidth]{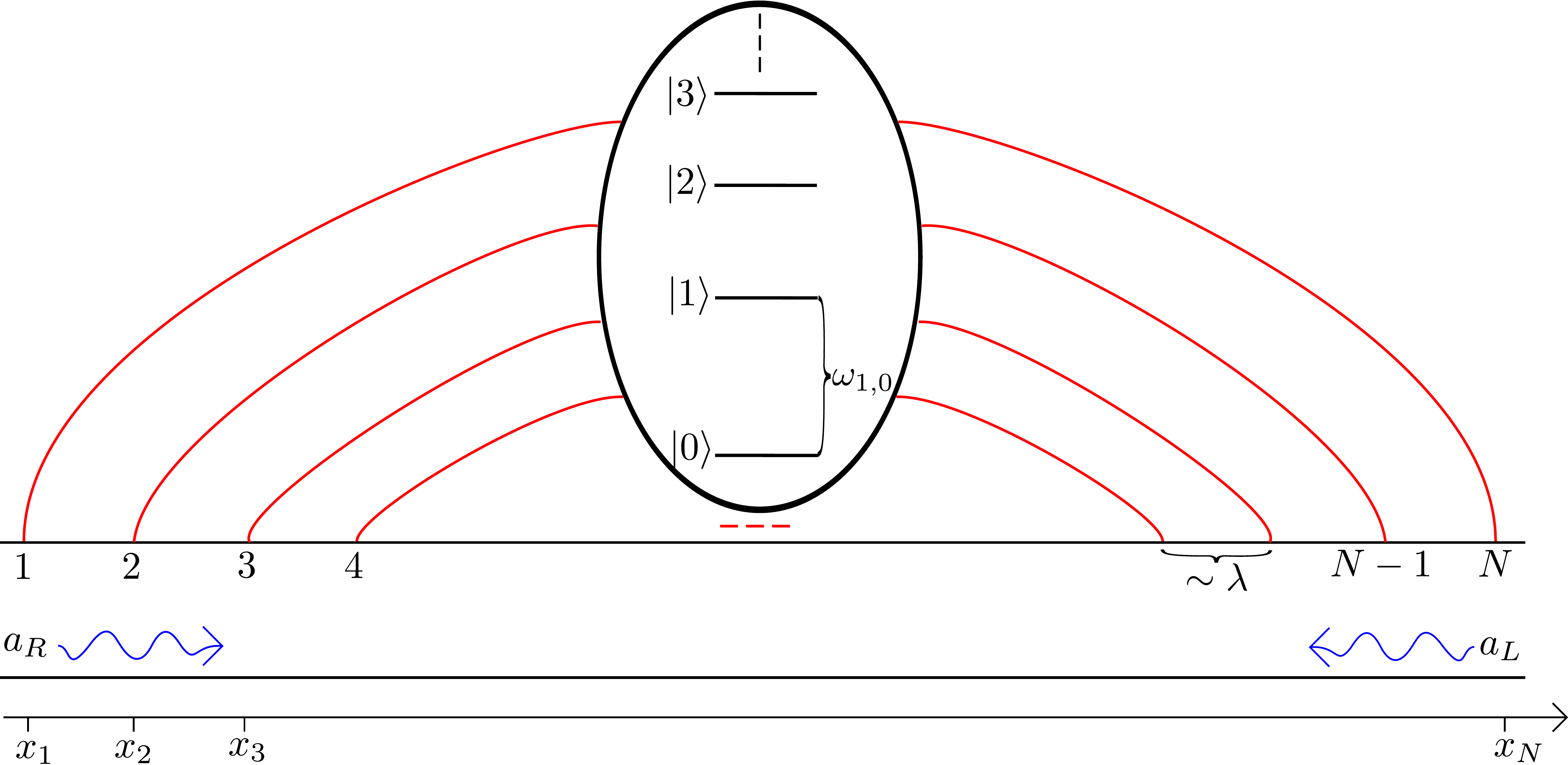}
\caption{A sketch of the system under consideration. A multilevel atom with energy levels $\ket{0}, \ket{1}, \ket{2}, \ldots$ couples at the points $x_1, \ldots, x_N$ to a bosonic field with right- and left-travelling modes. The distance between the coupling points can for example be on the order of wavelengths $\lambda = 2\pi v/\omega_{1,0}$, where $\omega_{1,0}$ is the first transition frequency of the atom and $v$ is the velocity of the bosonic modes.\label{FigSetupSketch}}
\end{figure*}

While there have been experiments \cite{AstafievScience2010,AbdumalikovPRL2010,HoiPRL2011,HoiPRL2012,HoiPRL2013,HoiNJP2013} and theoretical studies \cite{ShenPRL2005,PeropadreNJP2013} with an atom coupled at a single point to a one-dimensional (1D) field, and also with several atoms coupled to the field at different points \cite{Lalumiere2013,vanLooScience2013,LehmbergPRA1970,FicekPRA1990,LenzPRA1993,FicekPRep2002,OrdonezPRA2004,KienPRA2005,GonzalesPRL2011,ZuecoPRB2012,ZhengBarangerPRL2013}, to the best of our knowledge, the situation outlined above has not been studied previously. In this paper, we therefore investigate the physics of an atom coupled to a massless 1D bosonic field at several connection points, which may be spaced wavelengths apart.

When the atom couples to the field at a single point it interacts with vacuum fluctuations, leading to relaxation at its transition frequencies, and to a renormalization of those frequencies known as the Lamb shift \cite{CarmichaelBook1,GardinerZoller,LambPR1947,BethePR1947}, which has been studied also for superconducting qubits \cite{FragnerScience2008,GramichPRA2011}. Introducing more connection points opens up the possibility of interference playing a role in these processes. The result is that we can calculate the frequency-dependence of the atom coupling strength and Lamb shift for a given structure, or conversely design a certain frequency dependence by choosing the spacing between the connection points. Essentially, this is done by performing a discrete Fourier transform of the inter-point distances \cite{FollandFourierBook}, as the wave vector is related to the frequency via the boson velocity.

Classically, these interference effects are well-known for SAW systems in commercial use \cite{DattaSAWBook,MorganSAWBook}. Bringing them to the quantum world would be an interesting generalization of the spin boson model \cite{LeggettRMP1987,CaldeiraLeggettPRL1981}. While there have been papers investigating the effect of a few particular frequency-dependent couplings between atom and field \cite{LewensteinPRA1988,WilhelmChemPhys2004,QuangPRL1997,WangPRL2004}, there has, as far as we know, not been any previous study showing how couplings with arbitrary frequency dependencies can be realized in quantum optics. We note, however, that a precursor of these interference effects can be seen in studies of an atom placed in front of a mirror \cite{MeschedePRA1990,HindsPRA1991,EschnerNature2001,DornerPRA2002,KoshinoNJP2012,HoiInPreparation2014}, which lets the atom interact twice with the field.

Frequency-dependent couplings could be useful in a number of ways. Essentially, the applications are all based on changing the ratio between coupling strengths for transitions at different frequencies. For example, by changing the transition frequency of a qubit we could tune it from interacting strongly with the field to a frequency where the interaction is zero, thus protecting it from the environment. One can also imagine placing two transitions at very different coupling strengths to facilitate a population inversion needed for lasing \cite{YouPRB2007}, or amplifying multi-photon processes by tuning the frequencies of lower order processes to interaction minima.

This article is organized as follows. In \secref{SecTheSystem}, we describe the system. We sketch a derivation of the effective master equation for the atom, considering both the situation of an open transmission line and that of the atom being placed close to a mirror. Then, in \secref{SecFrequency}, we investigate the frequency dependence of the coupling strength between the atom and the environment and of the Lamb shift of the atom. We show that by controlling the coupling strength at each connection point and the distance between connection points, a wide variety of frequency dependencies can be designed for the total coupling. Some possible applications of such designed frequency-dependent couplings are then discussed in \secref{SecApplications}. The applications include tunable coupling, single-atom lasing and various two-tone experiments. In \secref{SecExperiments}, we discuss possible experimental realizations of our system. In \secref{SecConclusion}, finally, we conclude and give an outlook for future work.

The calculations referred to in \secref{SecTheSystem} are presented in detail in the appendices. In \appref{AppContinuumCalc}, we do the standard master equation derivation by tracing out the environment. Then, in \appref{AppSLHCalc}, we use the equivalent $(S,L,H)$ formalism for cascaded quantum systems to redo the calculations in a different way, and also to handle the case of the giant artificial atom placed in front of a mirror. 


\section{Giant atom}
\label{SecTheSystem}

\subsection{Hamiltonian}

The system we consider is sketched in \figref{FigSetupSketch}. A multi-level atom is connected at $N$ points to right- and left-moving modes of a bosonic field obeying the massless Klein-Gordon equation. The Hamiltonian of the system is given by
\be
H = H_A + H_F + H_I,
\label{TotalHamiltonian}
\ee
where we have defined the atom Hamiltonian
\bea
H_A &=& \sum_m \omega_m \ketbra{m}{m},
\eea
the field Hamiltonian
\bea
H_F &=& \sum_j \omega_j \left(a_{Rj}^\dag a_{Rj} + a_{Lj}^\dag a_{Lj}\right),
\eea
and the interaction Hamiltonian
\bea
H_I &=& \sum_{j,k,m} g_{jkm} \left(  \ketbra{m}{m+1} + \ketbra{m+1}{m} \right) \nn\\
&&\times\bigg(a_{Rj} e^{-i\omega_j x_k/v} + a_{Lj} e^{i\omega_j x_k/v} \nn\\
&&\quad+ a^\dag_{Rj} e^{i\omega_j x_k/v} + a^\dag_{Lj} e^{-i\omega_j x_k/v} \bigg),\:\:
\label{EqHI}  
\eea
respectively, all in units where $\hbar = 1$. The atom levels are labelled by the index $m = 0, 1, 2, \ldots$ and have energies $\omega_m$. The indices $R$ and $L$ denote right- and left-moving bosons, respectively, and the boson modes are furthermore distinguished by the index $j$. The coordinate of connection point $k$ is denoted $x_k$ and $v$ is the boson velocity, which we assume to be frequency-independent. We assume that the time it takes for a boson to travel between two connection points is negligible compared to the relevant timescales of the problem (the relaxation rate of the atom), leaving the phase shift $e^{i\omega_j x_k/v}$ as the only effect. In addition, we assume that the coupling strengths $g_{jkm}$ are small compared to the relevant $\omega_m$ and $\omega_j$ and that they can be factorized as $g_{jkm} = g_j g_k g_m$, which is the case for the transmon \cite{KochPRA2007}. In general, the mode coupling strength $g_j$ can be considered constant over a wide frequency range. The factors $g_k$ are dimensionless and only describe the relative coupling strengths of the different connection points. Finally, for the transmon \cite{KochPRA2007} and other atoms close to harmonic oscillators, we have $g_m = \sqrt{m+1}$.

\subsection{Master equation}
\label{SecTheSystemSubsecME}

In \appref{AppContinuumCalc}, we derive the master equation for the atom using standard techniques \cite{CarmichaelBook1,GardinerZoller}. We do not perform the rotating wave approximation (RWA) on the interaction Hamiltonian in \eqref{EqHI}, but do it on the master equation. This gives the correct expression for the Lamb shift \cite{AgarwalPRA1973,AckerhaltPRL1973,Agarwal1974}. Introducing the notation
\bea
\sm^m &=& \ketbra{m}{m+1}, \label{EqDefSm}\\
\sp^m &=& \ketbra{m+1}{m}, \\
\omega_{r,s} &=& \omega_{r} - \omega_{s}, \\
A(\omega_j) &=& g_j \sum_k g_k e^{i\omega_j x_k/v},
\label{EqDefA}
\eea
we arrive at the result
\bea
\dot{\rho}(t) &=& - i \comm{\sum_m \left(\omega_m + \Delta_m \right)\ketbra{m}{m}}{\rho(t)} \nn\\
&&+ \sum_m \Gamma_{m+1,m} \bigg[ \left(1 + \bar{n}(\omega_{m+1,m}) \right) \lind{\sm^m} \nn\\
&&\qquad\qquad\qquad\quad + \bar{n}(\omega_{m+1,m}) \lind{\sp^m} \bigg], \label{EqMasterEq}
\eea
where $\rho$ is the density matrix for the giant atom and we use the notation $\lind{X}\rho = X\rho X^\dag - \frac{1}{2} X^\dag X\rho - \frac{1}{2} \rho X^\dag X$ for the Lindblad superoperators \cite{Lindblad1976}.

Here, the relaxation rates $\Gamma_{m+1,m}$ for the transitions $\ket{m+1}\rightarrow\ket{m}$ are given by
\be
\Gamma_{m+1,m} = 4\pi g_m^2 J(\omega_{m+1,m}) \abssq{A(\omega_{m+1,m})}, \label{EqDefGammam}
\ee
where $J(\omega)$ is the density of states for the bosonic modes, and $\bar{n}(\omega,T)$ is the mean number of bosonic excitations at frequency $\omega$ and temperature $T$,
\be
\bar{n}(\omega,T) = \frac{e^{-\hbar\omega/k_B T}}{1 - e^{-\hbar\omega/k_B T}}. \label{EqDefn}
\ee
Furthermore, the energy shift $\Delta_m$ of level $m$ is 
\bea
\Delta_m &=& 2\mathcal{P} \int_0^\infty \id \omega J(\omega)\abssq{A(\omega)} \nn\\
&&\times \bigg( \frac{g_m^2\bar{n}(\omega,T)}{\omega-\omega_{m+1,m}} - \frac{g_m^2(1 + \bar{n}(\omega,T))}{\omega+\omega_{m+1,m}} \nn\\
&&\quad+ \frac{g_{m-1}^2 \bar{n}(\omega,T)}{\omega+\omega_{m,m-1}} - \frac{g_{m-1}^2 (1 + \bar{n}(\omega,T))}{\omega-\omega_{m,m-1}} \bigg). \label{EqDefDeltam}
\eea
where $\mathcal{P}$ denotes principal value (see \appref{AppContinuumCalc}). The terms without $\bar{n}(\omega,T)$ are the Lamb shift arising from interaction with the vacuum fluctuations of the bosonic field. The terms with $\bar{n}(\omega,T)$ are the Stark shift, which is due to interaction with thermal excitations of the field.

The difference compared to the case of a small atom is the frequency-dependent factor $\abssq{A(\omega)}$, which enters both in the expressions for the relaxation rate and for the Lamb shift. The expressions for a small atom would be recovered with $N=1$ and $\abssq{A(\omega)} = g_j^2$. In \secref{SecFrequency}, we explore the form of the frequency dependence that $\abssq{A(\omega)}$ gives rise to. 

For a 1D transmission line, we have the "ohmic" density of states $J(\omega)\propto \omega$. Limiting ourselves to the case of negligible temperature ($k_B T/\hbar\omega \ll 1$), we see that the expression for the Lamb shift would be diverging linearly for the case of a small atom. Renormalizing in the spirit of Bethe's calculation for the original Lamb shift \cite{BethePR1947}, we instead have (see \appref{AppContinuumCalc})
\bea
\Delta_m &=& 2\mathcal{P} \int_0^\infty \id \omega \frac{J(\omega)}{\omega}\abssq{A(\omega)} \nn\\
&&\times \bigg( \frac{g_m^2 \omega_{m+1,m}}{\omega+\omega_{m+1,m}} - \frac{g_{m-1}^2 \omega_{m,m-1}}{\omega-\omega_{m,m-1}} \bigg), \label{EqDefDeltamRenorm}
\eea
which still diverges for a small atom in a 1D transmission line, but only logarithmically. We can introduce a cutoff frequency $\omega_c$ to get a finite value. For a transmon with $\omega_{1,0} \approx \unit[5]{GHz}$ a reasonable choice for $\omega_c$ is the superconducting gap $\Delta_{SC} \approx \unit[100]{GHz}$, \emph{i.e.}, $\omega_c/\omega_{1,0} \approx 20$. For the case of a small 2-level atom, this would give a shift of the transition frequency by \cite{AgarwalPRA1973,AckerhaltPRL1973,Agarwal1974}
\be
\Delta_{1,0} = - \frac{\Gamma_{1,0}}{2\pi}\ln\left(\frac{\omega_c^2}{\omega_{1,0}^2} - 1\right) \approx 0.95 \Gamma_{1,0}.
\ee
For a small multi-level atom with weak anharmonicity, the shift of the transition frequencies is negligible. However, as we shall see in \secref{SecFrequency}, the result can be different for a giant atom both with two and more levels.

\subsection{(S,L,H) formalism and mirror}

An alternative way to derive the frequency dependence of the relaxation rates and the Lamb shifts is to use the $(S,L,H)$ formalism for cascaded quantum systems \cite{GoughCommMathPhys2009,GoughIEEE2009}. The underlying assumptions of that formalism are mostly the same as the ones we used above, $i.e.$, weak coupling and negligible travel time, but also include a constant density of states $J(\omega)$. We assume negligible temperature ($\bar{n}=0$) for simplicity.

The detailed $(S,L,H)$ calculations are shown in \appref{AppSLHCalc}. The result for a two-level atom is a relaxation rate
\be
\Gamma_{1,0} = \abssq{\sum_{k=1}^N \sqrt{\gamma_k}\exp\left(i\sum_{j=1}^{k-1} \phi_j \right)} \label{EqRelaxSLH}
\ee
and a Lamb shift
\be
\Delta_1 = \sum_{i=1}^{N-1}\sum_{k=1}^{N-i} \sqrt{\gamma_k \gamma_{k+i}} \sin\left(\sum_{j=k}^{k+i-1} \phi_j\right),
\label{EqLambSLH}
\ee
where the relaxation rate for a single connection point is $\gamma_k$ and the phase shift from one connection point to the next is written $\phi_k=\omega_{1,0}(x_{k+1}-x_k)/v$. The result for the relaxation rate is the same as \eqref{EqDefGammam} with $\bar{n}(\omega)=0$ and $J(\omega)=J(\omega_{1,0})$ inserted, since we can identify 
\be
\gamma_k = 4\pi g_j^2 g_k^2 J(\omega_{1,0}).
\ee
Similarly, the Lamb shift term in \eqref{EqLambSLH} is the result obtained for low temperature and constant density of states in \eqref{EqDefDeltamRenorm}, considering only the dominating second term and extending the lower limit to $-\infty$, \emph{i.e.},
\be
\Delta_1 = -  2 \mathcal{P}\int_{-\infty}^\infty \id \omega \frac{ J(\omega_{1,0}) \abssq{A(\omega)}}{\omega - \omega_{1,0}}. \label{EqLambHilbert}
\ee
This captures the contribution to the Lamb shift from the interaction at frequencies close to $\omega_{1,0}$.

An added benefit of doing the calculations in the $(S,L,H)$ formalism is that it becomes easy to treat the case where the giant atom is placed in front of a mirror. The result, derived in \appref{AppSLHCalc}, for the mirror to the right of the atom, is a modified relaxation rate
\be
\Gamma_{1,0}^{\text{mirror}} = \abssq{A_L(\left\{\gamma_j,\phi_j\right\}) + e^{i(\phi_\Sigma + \phi_M)}A_R(\left\{\gamma_j, \phi_j\right\})}
\ee
and an addition of $\im\left(A_R^2 e^{i\phi_M}\right)$ to the Lamb shift. Here, $\phi_M$ is the phase shift acquired during the travel to the mirror and back. We have assumed the corresponding travel time to be negligible just like the travel time across the giant atom. We have also used the notation
\bea
\phi_\Sigma &=& \sum_{j=1}^{N} \phi_j, \\
A_L(\left\{\gamma_k,\phi_k\right\}) &=& \sum_{k=1}^N \sqrt{\gamma_k/2}\exp\left(i\sum_{j=1}^{k-1} \phi_j \right), \\
A_R(\left\{\gamma_k,\phi_k\right\}) &=& \sum_{k=1}^N \sqrt{\gamma_k/2} \exp\left(i\sum_{j=k}^{N-1} \phi_j \right),
\eea
where $A_L$ and $A_R$ contain the phase factors for left- and right-moving bosons, respectively. We note that $\abssq{A_L(\left\{\gamma_k,\phi_k\right\})} = \abssq{A_R(\left\{\gamma_k,\phi_k\right\})}$ and
\be
\abssq{A_L(\left\{\gamma_k,\phi_k\right\})} + \abssq{A_R(\left\{\gamma_k,\phi_k\right\})} = \Gamma_{1,0}.
\ee
%


\section{Frequency-dependent coupling strength and Lamb shift}
\label{SecFrequency}

With the general expressions for the frequency-dependent relaxation rates and Lamb shifts given in Eqs.~(\ref{EqDefGammam}), (\ref{EqDefDeltamRenorm}), and (\ref{EqRelaxSLH})-(\ref{EqLambSLH}), we now turn our attention to the actual form of the frequency dependence.

\subsection{Maximally symmetric case}

We first consider the maximally symmetric case, where the coupling strength is the same at each connection point and the distance between neighbouring connection points is constant. This case is relevant for a recent experiment, coupling a transmon to surface acoustics waves \cite{GustafssonApr2014}. The symmetry implies that we can set $g_k=1$ in \eqref{EqDefA} or correspondingly $\gamma_k = \gamma$ in Eqs.~(\ref{EqRelaxSLH}) and (\ref{EqLambSLH}), and $\phi_k = \phi = \omega_{1,0}(x_2-x_1)/v$ in Eqs.~(\ref{EqRelaxSLH}) and (\ref{EqLambSLH}). The result from the $(S,L,H)$ expressions is a relaxation rate
\bea
\Gamma_{1,0}(\omega_{1,0}) = \gamma\frac{\sin^2\left(\frac{N}{2}\phi\right)}{\sin^2\left(\frac{1}{2}\phi\right)} = \gamma\frac{1-\cos(N\phi)}{1-\cos(\phi)}
\eea
and a contribution to the Lamb shift
\bea
\Delta_1 = \gamma\sum_{k=1}^N (N-k)\sin(k\phi) = \gamma\frac{N\sin(\phi) - \sin(N\phi)}{2\left(1 - \cos(\phi)\right)}. \:\:\:\:\:\:\:\: \label{EqDelta1SLH}
\eea
The ground state is not shifted, so $\Delta_1 = \Delta_{1,0}$. Note that the result for a small atom with a single connection point would be $\Gamma_{1,0} = \gamma$ and $\Delta_0=\Delta_1 = 0$. For a small atom, the main part of the Lamb shift is due to a sum of contributions from a wide range of frequencies. With an increasing number of connection points in the giant atom, the dominant contribution to the Lamb shift is instead due to interaction at frequencies close to $\omega_{1,0}$ and this is captured by \eqref{EqDelta1SLH}.

\begin{figure}[t!]
\includegraphics[width=\columnwidth]{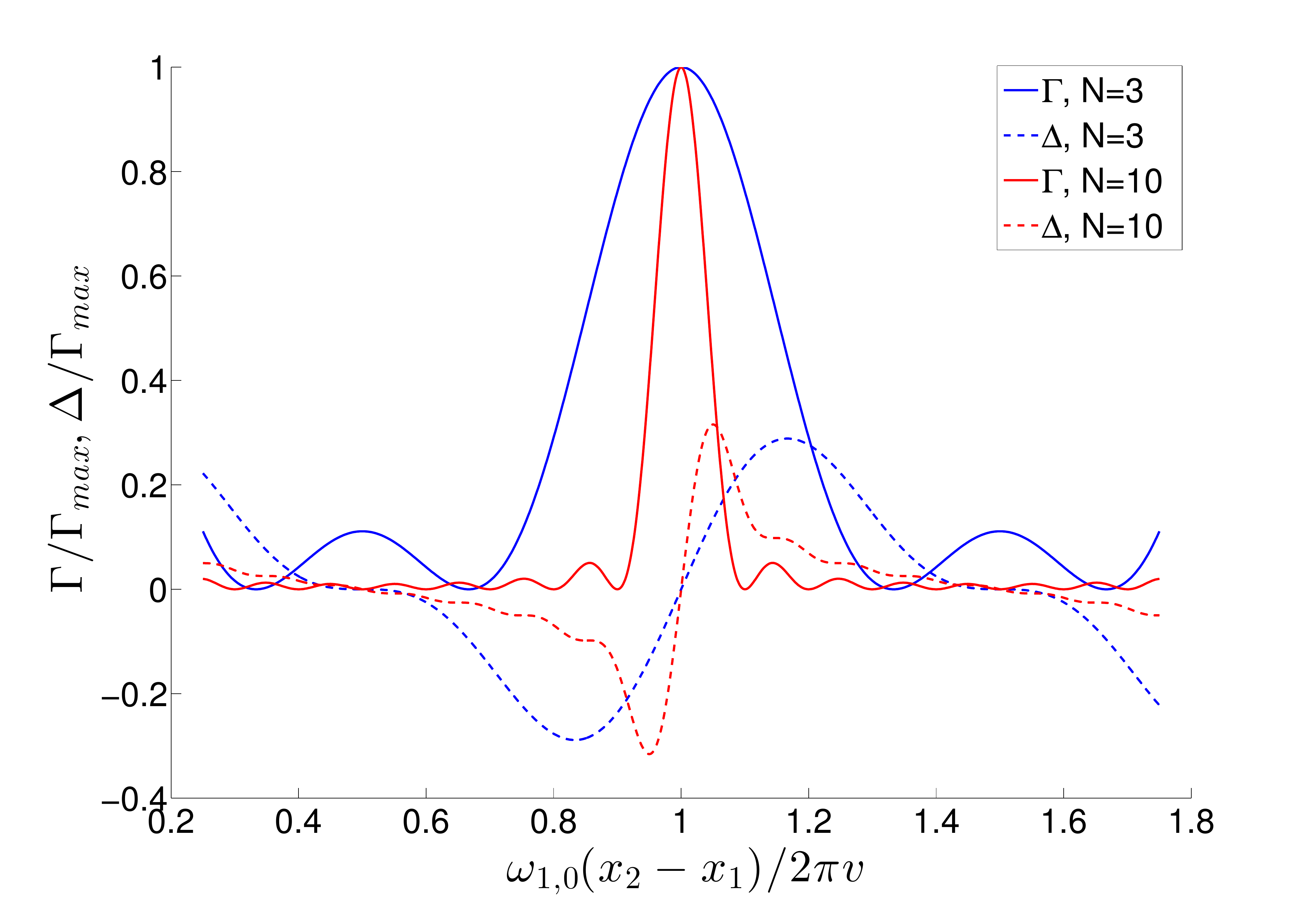}
\caption{The frequency dependence of the relaxation rate (solid lines) and main contribution to the Lamb shift (dashed lines) for $N=3$ (blue lines) and $N=10$ (red lines) in the symmetric case. Note that $\omega_{1,0}(x_2-x_1)/2\pi v$ corresponds to $\phi/2\pi$. Everything has been normalized to the maximum coupling strength for each N. We have set $J(\omega)$ constant for simplicity. It is usually a function varying slowly with $\omega$; in the "ohmic" case $J(\omega)\propto \omega$. \label{FigSymmetricN3N10}}
\end{figure}

We plot these results for the cases $N=3$ and $N=10$ in \figref{FigSymmetricN3N10}. For the relaxation rate, there is a clear maximum when the distance between neighbouring connection points correspond to an integer number $n$ of wavelengths for the transition frequency, \emph{i.e.}, $\omega_{1,0} = 2n\pi(x_2-x_1)/v$. There are also a number of lower, local maxima, but more interestingly we have a number of points where the relaxation rate goes to zero. This occurs when the distance between connection points is such that we get destructive interference in the coupling. The distance between maxima for the relaxation rate scales with $1/N$; more connection points give narrower resonances. The height of the global maximum scales with $N^2$.

For the contribution to the Lamb shift, we see that it can be both positive and negative. It is zero when the relaxation is maximum and it reaches its highest magnitude halfway between the relaxation maximum and the first relaxation minima. The Lamb shift is half the Hilbert transform of the relaxation rate, as shown in \eqref{EqLambHilbert}.

If we include the mirror close to the atom, we get in the symmetric case, with $\phi_m = \phi$, a relaxation rate
\be
\Gamma_{1,0}^{\text{mirror}}(\omega_{1,0}) = \frac{1}{2}\abssq{1+e^{iN\phi}}\Gamma_{1,0} = \gamma\frac{\sin^2(N\phi)}{2\sin^2\left(\frac{\phi}{2}\right)}
\ee
and a Lamb shift
\bea
\Delta_1^{\text{mirror}} &=& \Delta_1 + \frac{1}{2}\sin(N\phi)\Gamma_{1,0} \nn\\
&=& \gamma\frac{2N\sin(\phi) - \sin(2N\phi)}{4\left(1 - \cos(\phi) \right)}.
\eea
Effectively, the mirror lets the atom interact twice with the field and the result is that the frequency dependence of the relaxation rate and the Lamb shift gets twice the magnitude and twice as narrow resonances compared to the case without mirror.

\subsection{Designing the frequency dependence}

Moving on from the maximally symmetric case, we now ask ourselves what frequency dependencies we can create for the relaxation rates and the Lamb shifts given complete control over the coupling strength at each point and the spacing between connection points. The frequency dependence is determined by the $\abssq{A(\omega)}$, with $A(\omega)$ defined in \eqref{EqDefA}. We see that this is a discrete Fourier transform \cite{FollandFourierBook} of the coupling strengths at the individual connection points. Thus, given enough connection points and sufficient parameter control, in principle any frequency dependence of the relaxation rates can be designed.

\begin{figure}[t!]
\includegraphics[width=\columnwidth]{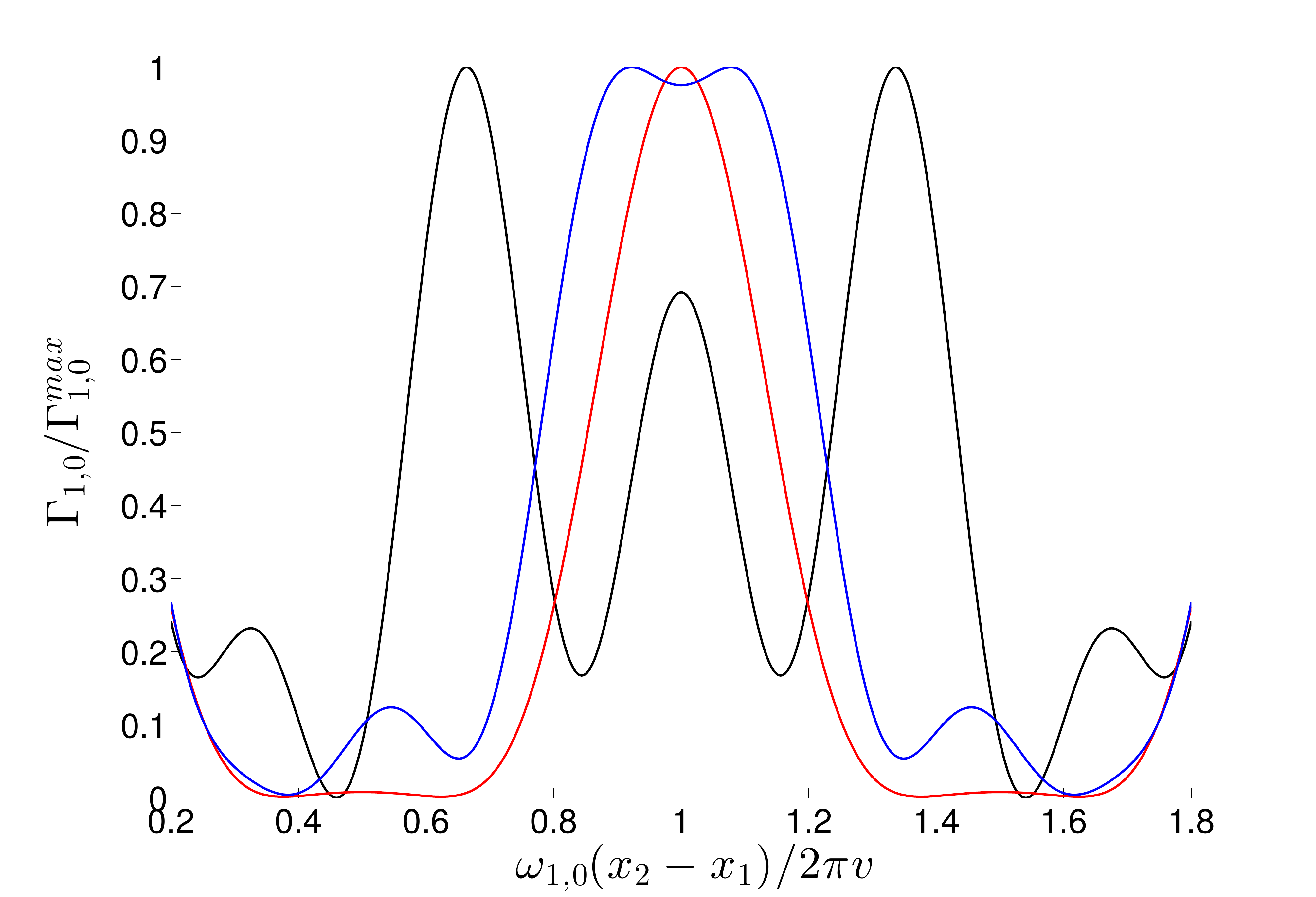}
\caption{Designed relaxation rate frequency dependencies. The black line shows two maxima of equal magnitude (parameters: $g_k = \{1,1,1,1\}$, $x_k = \{0,1,1.5,3\}x_2$), the blue line has a wide, flat maximum (parameters: $g_k = \{1,3,3,1\}$, $x_k = \{0,1,2,3.5\}x_2$), and the red line has two wide, shallow minima (parameters: $g_k = \{1,4,4,1\}$, $x_k = \{0,1,2,3\}x_2$). \label{FigDesignedFreqDeps}}
\end{figure}

To show just a few examples, in \figref{FigDesignedFreqDeps} we plot relaxation rates that have been tailored to have two maxima of equal magnitude (black line), a wide maximum (blue line), and wide, shallow minima (red line). This was done using only four connection points and just tuning a few parameters away from the maximally symmetric case.

\section{Applications}
\label{SecApplications}

In this section, we will discuss a number of possible applications for frequency-dependent relaxation rates and Lamb shifts. While there are several applications for the relaxation rates, it is harder to find a good use for the small Lamb shfts.

\subsection{Tunable coupling}

The ability to tune the coupling of an artificial atom to its surroundings is a desirable feature in many quantum information applications and has been realized for a transmon \cite{SrinivasanPRL2011}. Tunable coupling can limit interaction with the atom to only when it is needed for readout or control, leaving the atom protected from decoherence the rest of the time. Here, we see that a giant artificial atom can switch from strong coupling to the environment (a maximum in \figref{FigSymmetricN3N10}) to very weak coupling (a minimum in \figref{FigSymmetricN3N10}) by only changing the transition frequency slightly. For an artificial atom such as a transmon, it is easy to change the transition frequencies by controlling the magnetic flux through a SQUID loop. In fact, tunable coupling in this manner was demonstrated recently with a small artificial atom in front of a mirror (close to the case of $N=2$ for a giant artificial atom) in \cite{HoiInPreparation2014}. Ideally, it would perhaps be preferable to change the connection point distances \emph{in situ} rather than the transition frequency, but this seems hard to implement.

\subsection{Population inversion}
\label{SubSecLaser}

\begin{figure}[t!]
\includegraphics[width=\columnwidth]{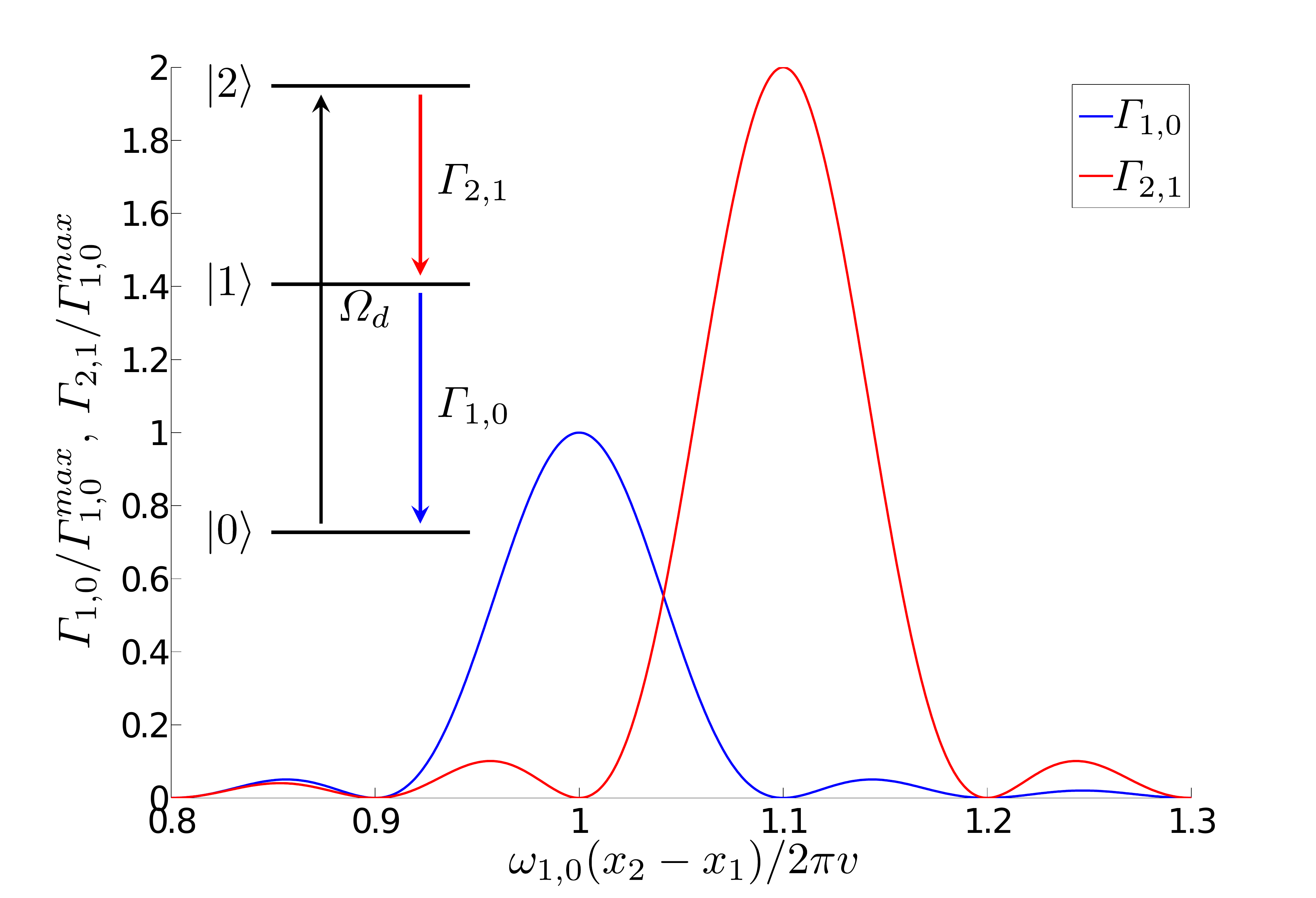}
\caption{A scheme for population inversion. The relaxation rates $\Gamma_{1,0}$ and $\Gamma_{2,1}$ for the first two atom transitions, plotted as a function of the first transition frequency $\omega_{1,0}$ for $N=10$ in the maximally symmetric case. By choosing the anharmonicity to be $-0.1\cdot 2\pi v/(x_2-x_1)$, we can make the global maximum of $\Gamma_{2,1}$ coincide with a minimum for $\Gamma_{1,0}$. Inset: Energy level diagram showing the relevant driving and relaxation rates for population inversion. \label{FigGamma1Gamma2InsetLevels}}
\end{figure}

Another application of the frequency-dependent relaxation rates involves higher levels of the atom. For the maximally symmetric case, we can have the situation depicted in \figref{FigGamma1Gamma2InsetLevels}. There we plot the relaxation rates $\Gamma_{1,0}$ and $\Gamma_{2,1}$ for an anharmonicity chosen in relation to $N$ such that $\Gamma_{2,1}$ has its global maximum when $\Gamma_{1,0}$ is at a minimum (and vice versa). This case opens up the possibility of lasing, as illustrated in the inset of \figref{FigGamma1Gamma2InsetLevels}. If we can drive the $\ket{0}\rightarrow\ket{2}$ transition with sufficient strength $\Omega_d$, we can achieve a population inversion. The giant atom will decay rapidly form $\ket{2}$ to $\ket{1}$, but the decay from $\ket{1}$ to $\ket{0}$ will be slow. 

Since the two decay rates can be very different, very strong population inversion should be obtainable. Placing the whole structure in a cavity should then allow to build a single-atom laser.

\subsection{Further possible applications}

\begin{figure}[t!]
\includegraphics[width=\columnwidth]{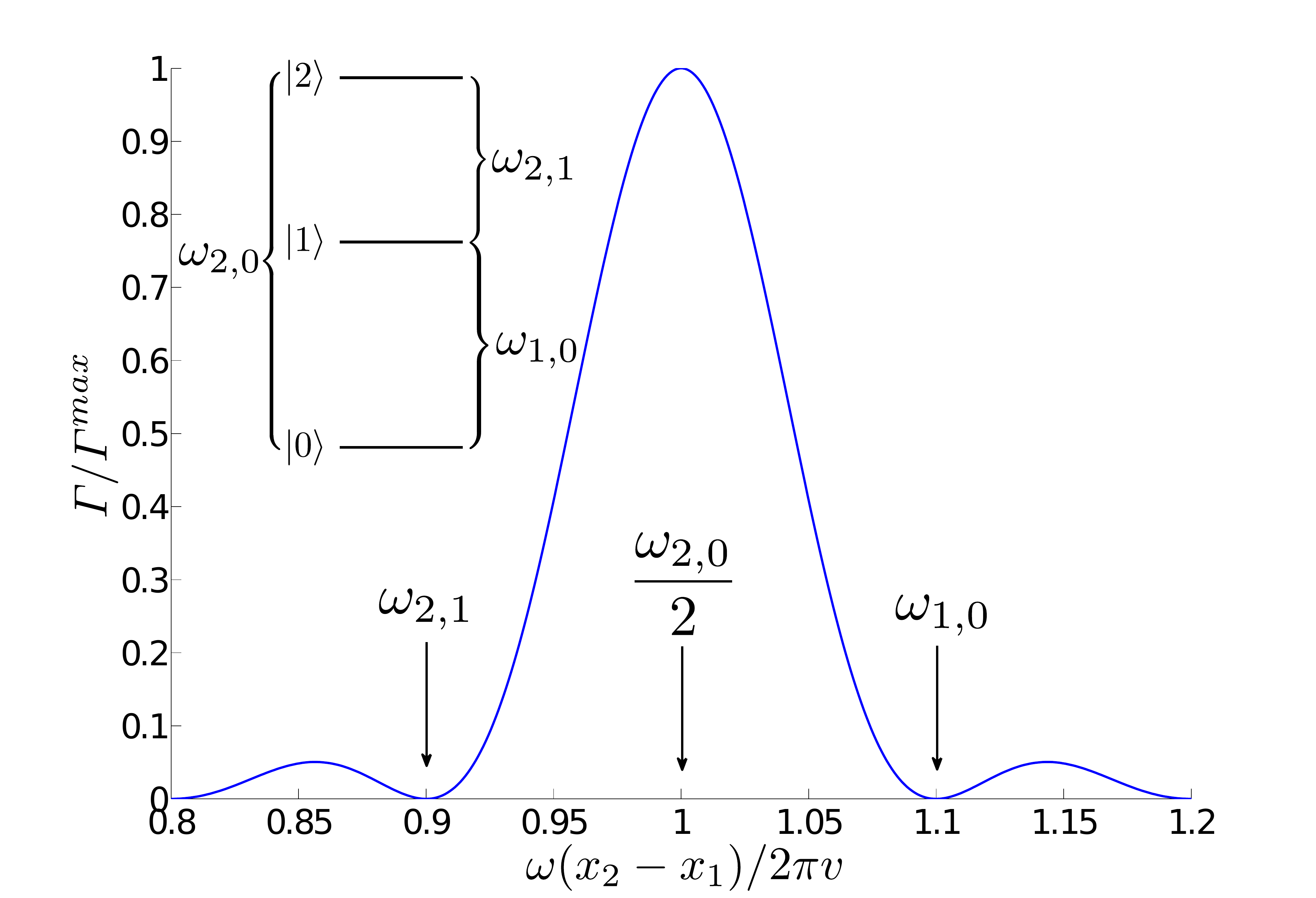}
\caption{Enhancing multi-photon relaxation rates. We plot relaxation rate as a function of frequency for the maximally symmetric case with $N=10$ and an anharmonicity $-0.2\cdot 2\pi v/(x_2-x_1)$. The $\ket{1} \rightarrow \ket{0}$ and $\ket{2} \rightarrow \ket{1}$ transitions can then be placed at relaxation rate minima while the two-photon process at $\omega_{2,0}/2 = (\omega_{1,0} + \omega_{2,1})/2$ is at a maximum. Inset: Energy level diagram showing the transition frequencies. \label{FigMultiPhoton}}
\end{figure}

There has been a few studies investigating specifically shaped environment structures $J(\omega)$ \cite{LewensteinPRA1988,WilhelmChemPhys2004}. Here, we can tailor $\abssq{A(\omega)}$ to achieve the same effect. Viewing the connection points as part of the atom, we can say that the atom "provides its own cavity", screening it from interacting with some modes. Building on this, a possible extension of the idea in \secref{SubSecLaser} would be to enhance multi-photon transition rates. One can easily imagine constructing a frequency-dependent relaxation rate with minima at single-photon transition frequencies and a maximum at some multi-photon transition frequency. To be explicit, consider for example the situation in \figref{FigMultiPhoton}, which can be arranged for an anharmonicity of  $-0.2\cdot 2\pi v/(x_2-x_1)$. The relaxation rates for the $\ket{1} \rightarrow \ket{0}$ and $\ket{2} \rightarrow \ket{1}$ transitions are both at minima when $\omega_{1,0}=1.1\cdot 2\pi v/(x_2-x_1)$, while the two-photon relaxation at frequency $\omega_{2,0}/2 = (\omega_{1,0} + \omega_{2,1})/2$ is at its maximum.

Another interesting subject to study both experimentally and theoretically would be the structure of the Autler-Townes doublet \cite{AutlerTownesPR1955,LewensteinPRA1988}, the splitting of the $\ket{0} \rightarrow \ket{1}$ transition into two due to a drive on the $\ket{1} \rightarrow \ket{2}$ transition, or the Mollow triplet \cite{MollowPR1969}, the splitting of the $\ket{0} \rightarrow \ket{1}$ transition into three due to a drive on the $\ket{0} \rightarrow \ket{1}$ transition, in a setting with frequency-dependent coupling.

\begin{figure}[t!]
\includegraphics[width=\columnwidth]{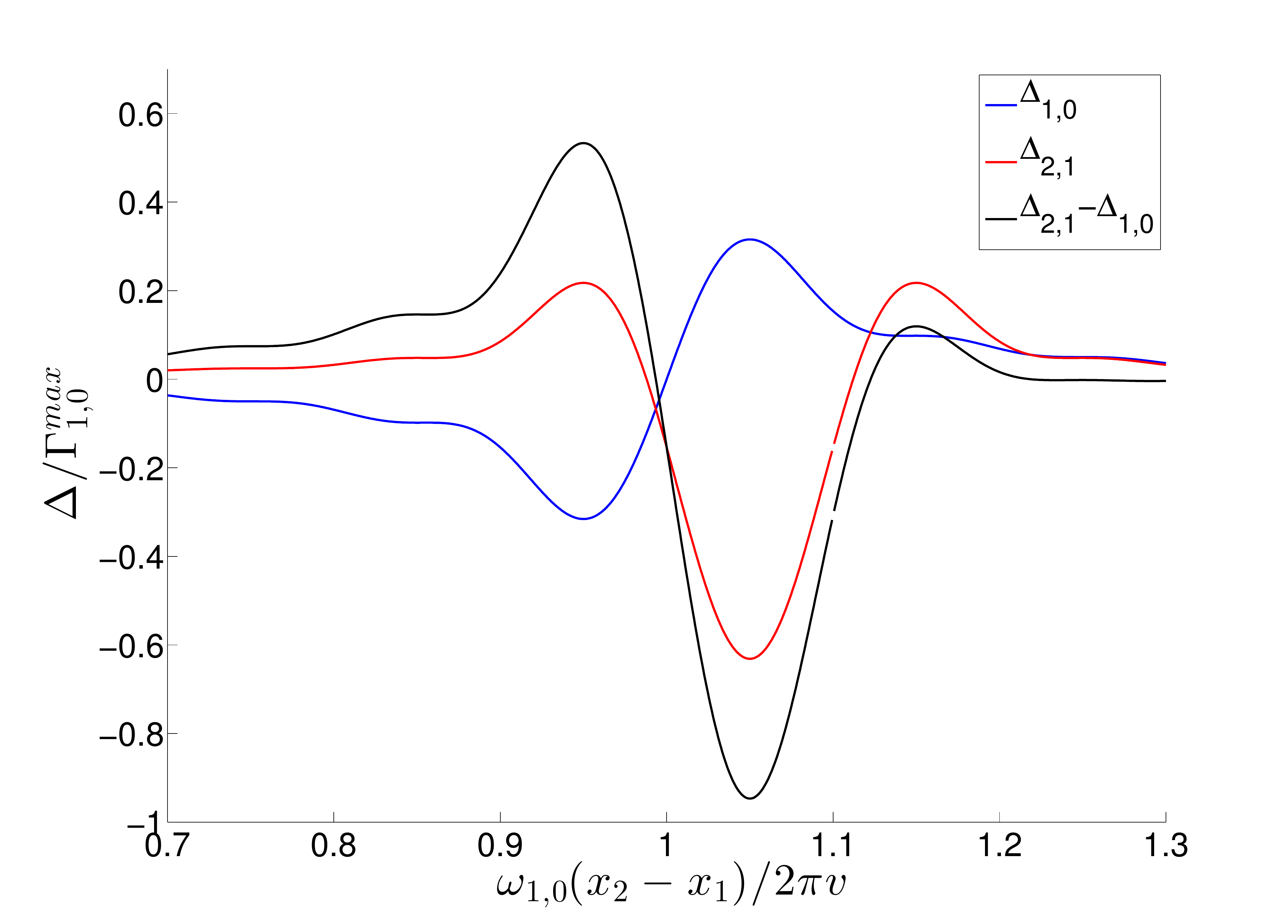}
\caption{Varying the anharmonicity. The Lamb shifts of the first (blue) and second (red) transitions of the giant atom plotted together with their difference (black), the resulting change in anharmonicity, for the maximally symmetric case with $N=10$ and an anharmonicity of $-0.1\cdot 2\pi v/(x_2-x_1)$. \label{FigAnharm}}
\end{figure}

Finally, it should also be possible to engineer a varying anharmonicity. Remember from \figref{FigSymmetricN3N10} that the Lamb shift changes sign on resonance in the maximally symmetric case. Positioning the $\ket{0} \rightarrow \ket{1}$ and  $\ket{1} \rightarrow \ket{2}$ transition frequencies on either side of the resonance would thus change the anharmonicity. This is illustrated in \figref{FigAnharm}. Note that we have assumed the anharmonicity to be much larger than the relaxation rate when deriving the master equation in \secref{SecTheSystem}. This means that the variation in the anharmonicity cannot be large compared to the total anharmonicity.

\section{Possible experimental implementations}
\label{SecExperiments}

Giant atoms with a number of discrete connection points are not readily available in nature, but there seem to be at least two straightforward ways to implement our system using artificial atoms made out of superconducting circuits.

\subsection{Transmon coupled to SAW}

\begin{figure}[t!]
\includegraphics[width=\columnwidth]{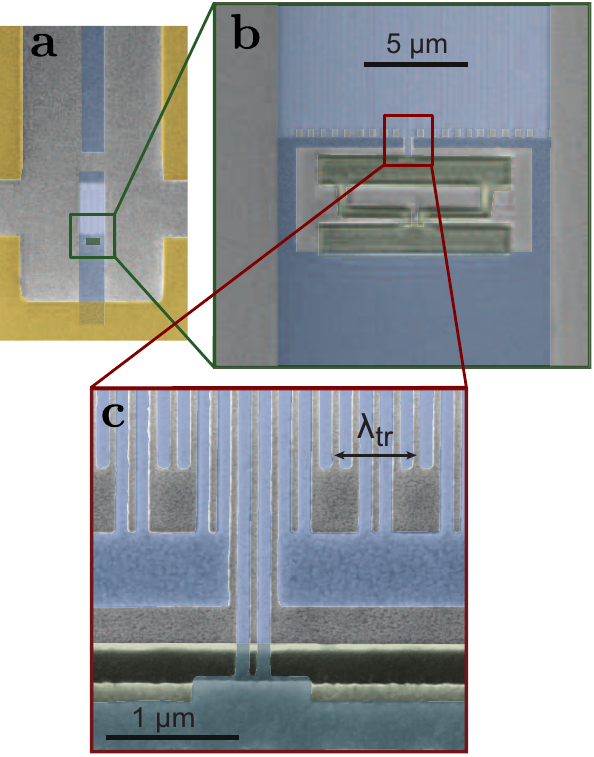}
\caption{An example of an experimental implementation of our system, using a transmon coupled to SAWs. Figure adapted from \cite{GustafssonApr2014} with thanks to M. V. Gustafsson, T. Aref, and M. K. Ekstr\"om for providing the images. a) The lower blue part is the two transmon islands. SAWs propagate from left to right in the gap between the grounded yellow areas. The upper blue part is an electrical gate, enabling RF excitation of the transmon. b) Zoom-in on the transmon islands. The green part is the SQUID connecting the islands. c) Zoom-in on the individual fingers of the transmon capacitance. The distance between neighbouring fingers (connection points) is on the order of the SAW wavelength. The double-finger structure used here reduces mechanical reflections. \label{FigTransmonSAW}}
\end{figure}

The first implementation, which motivated this work, was suggested in \cite{GustafssonNatPhys2012} and realized in \cite{GustafssonApr2014}. Here, the giant artificial atom is a transmon \cite{KochPRA2007}. It is not coupled to propagating microwave photons, as is the usual case, but it interacts instead with phonons in the form of surface acoustic waves (SAWs) \cite{DattaSAWBook,MorganSAWBook} propagating on a piezoelectric substrate. The setup is illustrated in \figref{FigTransmonSAW}. 

The interdigitated capacitance between the two islands of the transmon forms a transducer which couples to the SAWs. Due to the low SAW velocity, the distance between neighbouring fingers is on the order of wavelengths ($\lambda \approx \unit[10^{-6}]{m}$), realizing the necessary conditions for the physics described in this paper. A large number of connection points can easily be implemented.

From classical SAW theory \cite{DattaSAWBook,MorganSAWBook} we know that there are a number of transducer configurations possible, which could implement particular frequency dependencies for the relaxation rates of the transmon. Although the transition frequency of the transmon is a few GHz, which is higher than most industrial applications for SAWs, it should still be possible to achieve the lithographic precision needed to fine-tune distances between coupling points. To tune the coupling strength for a connection point, one could add a thin layer of nonpiezoelectric material between the piezoelectric substrate and the electrode finger of the transmon. The thickness of this layer could be varied between fingers to achieve varying coupling strengths. 

Finally, we note that it is not clear for which finger widths the approximation of point-like connection points remains valid. 

\subsection{Transmon coupled to meandering transmission line}

\begin{figure}[t!]
\includegraphics[width=\columnwidth]{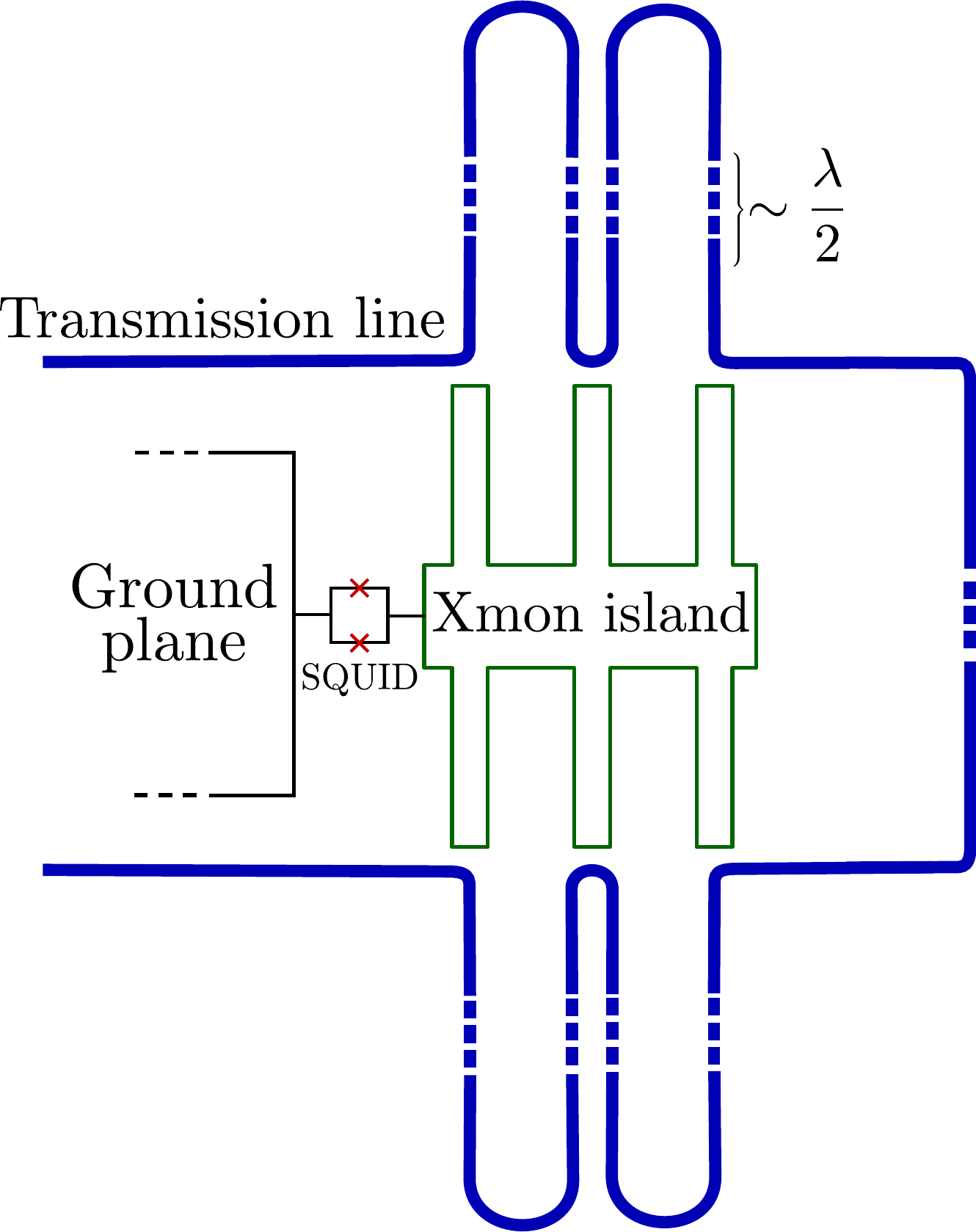}
\caption{A sketch of a possible implementation using an xmon coupled to a meandering transmission line. The distance between coupling points can be set with great precision by choosing the transmission line length and the capacitive coupling at each connection point can be tuned by designing the tips of the fingers of the xmon island. \label{FigXmonTL}}
\end{figure}

The second possible implementation of a giant artificial atom we foresee also uses a transmon. To be specific, it is a variation of the transmon known as the xmon \cite{BarendsPRL2013}, and it couples to an ordinary microwave transmission line. The intended setup is sketched in \figref{FigXmonTL}. 

The capacitive coupling between the transmission line and a finger of the xmon island can be designed with good accuracy, making possible large variations in relative coupling strengths between connection points. Furthermore, the distance from one connection point to the next can be made to be on the order of wavelengths by meandering the transmission line to fit it on a chip. This should give great precision in the control of the phase shifts between connection points. The drawback compared to the implementation with SAW is the size of the system. It will likely be hard to fit hundreds of wavelengths worth of transmission line on a single chip to investigate very large values of $N$ or connection point distances.


\section{Conclusion and outlook}
\label{SecConclusion}

We have studied the physics of an atom coupled to a 1D bosonic field at several connection points. The connection points can be spaced far apart, making the atom large compared to the wavelength of the field, an unusual situation which only recently has been realized in an experiment \cite{GustafssonApr2014}. We find that both the strength of the coupling and the size of the Lamb shift of the atom become frequency-dependent and that the dependence is determined by the discrete Fourier transform of the connection point coordinates.

We have discussed two possible experimental implementations of the system studied here. One is to couple a transmon to SAWs, another to couple it to a meandering microwave transmission line. In both cases, we can choose the coordinates of the connection points with great precision, thus enabling the design of a desired frequency dependence of the coupling strength. Since we can design the couplings this way, we can tune the ratio between the coupling strengths for transitions at different frequencies. We show here that this can be used to achieve tunable coupling, single-atom lasing, and amplification of multi-photon processes. Other applications can probably be found by comparison with classical SAW theory, which has been widely used for different kinds of filters for several decades \cite{DattaSAWBook,MorganSAWBook}.

In this work, we assumed that the relaxation time of the atom was much longer than all other relevant timescales, including the time it takes to travel from the first connection point to the last. An interesting direction for future work is to relax this assumption and investigate what happens when the travel time is not negligible. This is reminiscent of an atom placed far from a mirror, which has been studied before \cite{DornerPRA2002}, and should also connect to recent work on two atoms placed far apart \cite{ZhengBarangerPRL2013}. In particular, one could investigate the physics of the atom interacting with a pulse which is shorter than the travel time across the atom. In light of the recent interest in and progress on the topic of ultra-strong coupling \cite{BraakPRL2011,BourassaPRA2012,PeropadrePRL2013,NiemczykNatPhys2010,FornDiazPRL2010,ScalariScience2013,XiaPRB2011}, it would also be interesting to see what happens when the coupling at a single connection point, or the total coupling, becomes non-negligible compared to the atom frequency. Other possibilities for future work include placing the giant atom in a cavity and relaxing the assumption that signals travel instantaneously from the connection point to the atom.  

\section{Acknowledgments}\label{sec_ack}

We thank Martin Gustafsson, Vitaly Shumeiko, Thomas Aref, Lars Tornberg, Juan José García-Ripoll, Tom Stace and Gerard Milburn for valuable discussions. We acknowledge financial support from the Swedish Research Council and from the EU through the ERC and the project ScaleQIT.

\bibliographystyle{apsrev4-1}
\bibliography{BibRefs}


\appendix

\section{Detailed derivation of the master equation}
\label{AppContinuumCalc}

In this appendix, we perform the full derivation of the master equation given in \secref{SecTheSystemSubsecME}. We follow the standard procedure for tracing out the environment as given in Refs.~\cite{CarmichaelBook1,GardinerZoller}.

The Hamiltonian is given in Eqs.~(\ref{TotalHamiltonian})-(\ref{EqHI}). Moving to the interaction picture by transforming all operators according to
\be
\tilde{X}(t) = e^{i (H_A + H_F) t} X e^{-i (H_A + H_F) t},
\ee
we have the master equation
\be
\dot{\tilde{\rho}}_\text{tot}(t) = -i \comm{\tilde{H}_I(t)}{\tilde{\rho}_\text{tot}(t)},
\ee
where $\rho_\text{tot}$ is the density matrix of field and atom together. Integrating this equation once, reinserting the result and then tracing over the field degrees of freedom gives
\bea
\dot{\tilde{\rho}}(t) &=& \text{Tr}_F \bigg( -i \comm{\tilde{H}_I(t)}{\tilde{\rho}_\text{tot}(0)} \nn\\
&&\qquad\:\: - \int_0^t \id \tau \comm{\tilde{H}_I(t)}{\comm{\tilde{H}_I(\tau)}{\tilde{\rho}_\text{tot}(\tau)}} \bigg). \quad
\label{EqExactRhoBeforeApprox}
\eea
We now make the Born approximation, assuming the coupling between the field and the atom to be weak enough, and the "bath" provided by the field large enough, that the field remains in a thermal equilibrium state: $\rho_F(t) = \rho_F$. Furthermore, we make the Markov approximation that bath correlations decay rapidly compared to the timescale of the atom evolution, so $\dot{\rho}(t)$ can only be a function of $\rho(t)$. Finally also assuming the field and the atom to be uncorrelated at time $t=0$, \eqref{EqExactRhoBeforeApprox} reduces to
\be
\dot{\tilde{\rho}}(t) = - \int_0^t \id \tau \text{Tr}_F \left(  \comm{\tilde{H}_I(t)}{\comm{\tilde{H}_I(\tau)}{\tilde{\rho}(t)\rho_F}} \right).
\label{EqRhoqHIIntegral}
\ee

For brevity, the interaction Hamiltonian is written in terms of atom operators $s$ and bath operators $b$,
\be
H_I = sb + s b^\dag + s^\dag b + s^\dag b^\dag.
\ee
In the interaction picture, we identify
\bea
\tilde{s}(t) &=& \sum_m g_m \smm e^{-i\omega_{m+1,m}t}, \\
\tilde{b}(t) &=& \sum_j \left(a_{Rj} A^\dag(\omega_j) + a_{Lj}A(\omega_j) \right) e^{-i\omega_j t}, \label{EqDefTildeb}
\eea
where we have used the definitions from Eqs.~(\ref{EqDefSm})-(\ref{EqDefA}). Inserting this into \eqref{EqRhoqHIIntegral}, we apply the RWA to eliminate all rapidly rotating terms with $ss$ and $s^\dag s^\dag$. Using the notation $\expec{AB}_F = \text{Tr}_F(AB\rho_F)$ and noting that $\expec{bb}_F = \expec{b^\dag b^\dag}_F = 0$ we arrive at
\begin{widetext}
\bea
\dot{\tilde{\rho}}(t) &=&  - \int_0^t \id \tau \bigg[ \left(\expec{\tilde{b}(t)\tilde{b}^\dag(\tau)} + \expec{\tilde{b}^\dag(t)\tilde{b}(\tau)} \right) \left(\tilde{s}(t)\tilde{s}^\dag(\tau)\rho + \tilde{s}^\dag(t)\tilde{s}(\tau)\rho - \tilde{s}(\tau)\rho\tilde{s}^\dag(t) - \tilde{s}^\dag(\tau)\rho\tilde{s}(t) \right) \nn\\
&& + \left(\expec{\tilde{b}(\tau)\tilde{b}^\dag(t)} + \expec{\tilde{b}^\dag(\tau)\tilde{b}(t)} \right) \left(\rho\tilde{s}(\tau)\tilde{s}^\dag(t) + \rho\tilde{s}^\dag(\tau)\tilde{s}(t) - \tilde{s}(t)\rho\tilde{s}^\dag(\tau) - \tilde{s}^\dag(t)\rho\tilde{s}(\tau) \right) \bigg].
\label{EqRhoqHIIntegralExpec}
\eea
From \eqref{EqDefTildeb} we calculate
\bea
\expec{b(t)b^\dag(\tau)}_F &=& 2 \sum_j \abssq{A(\omega_j)}\left(1 + \bar{n}(\omega_j,T)\right) e^{-i\omega_j(t-\tau)}, \\
\expec{b(\tau)b^\dag(t)}_F &=& 2 \sum_j \abssq{A(\omega_j)}\left(1 + \bar{n}(\omega_j,T)\right) e^{i\omega_j(t-\tau)}, \\
\expec{b^\dag(t)b(\tau)}_F &=& 2 \sum_j \abssq{A(\omega_j)} \bar{n}(\omega_j,T) e^{i\omega_j(t-\tau)}, \\
\expec{b^\dag(\tau)b(t)}_F &=& 2 \sum_j \abssq{A(\omega_j)} \bar{n}(\omega_j,T) e^{-i\omega_j(t-\tau)},
\eea
where $\bar{n}(\omega_j,T)$ is the number of excitations in mode $j$ at temperature $T$ as defined in \eqref{EqDefn}. Inserting these results into \eqref{EqRhoqHIIntegralExpec}, using the full expressions for $s$ and $b$ gives
\bea
\dot{\tilde{\rho}}(t) &=& - 2\sum_{j,m} g_m^2 \abssq{A(\omega_j)} \int_0^t \id \tau \bigg[ \left( \left(1 + \bar{n}(\omega_j,T)\right) e^{-i\omega_j(t-\tau)} + \bar{n}(\omega_j,T) e^{i\omega_j(t-\tau)} \right) \nn\\
&&\times \left(\smm\spm\rho e^{-i\omega_{m+1,m}(t-\tau)} + \spm\smm\rho e^{i\omega_{m+1,m}(t-\tau)} - \smm\rho\spm e^{i\omega_{m+1,m}(t-\tau)} - \spm\rho\smm\rho e^{-i\omega_{m+1,m}(t-\tau)}\right) \nn\\
&&+ \left( \left(1 + \bar{n}(\omega_j,T)\right) e^{i\omega_j(t-\tau)} + \bar{n}(\omega_j,T) e^{-i\omega_j(t-\tau)} \right) \nn\\
&&\times \left(\rho\smm\spm e^{i\omega_{m+1,m}(t-\tau)} + \rho\spm\smm e^{-i\omega_{m+1,m}(t-\tau)} - \smm\rho\spm e^{-i\omega_{m+1,m}(t-\tau)} - \spm\rho\smm e^{i\omega_{m+1,m}(t-\tau)} \right) \bigg].\quad
\eea
Here we have assumed the anharmonicity of the atom to be large compared to the inverse timescale of the atom relaxation, allowing us to use the RWA to eliminate terms containing $\smm$ and $\sp^{m'}$ with $m\neq m'$. We now make the change of variables $t' = t-\tau$. Since we are interested in timescales $t\gg 1/\omega_{m+1,m}$, we can extend the upper integration limit in the $t'$ integral to infinity. We also replace the sum over $j$ with an integral over $\omega$, including the density of states $J(\omega)$, giving
\bea
\dot{\tilde{\rho}}(t) &=& 2\sum_{m} g_m^2 \int_0^\infty \id\omega J(\omega) \abssq{A(\omega)} \int_0^\infty \id t' \nn\\
&& \times \bigg[ e^{-i(-\omega + \omega_{m+1,m}) t'} \left\{  \bar{n}(\omega,T) \left(- \smm\spm\rho + \spm\rho\smm \right) + \left(1 + \bar{n}(\omega,T)\right)\left( - \rho\spm\smm + \smm\rho\spm \right) \right\} \nn\\
&&+ e^{-i(-\omega -\omega_{m+1,m}) t'} \left\{  \bar{n}(\omega,T) \left(- \spm\smm\rho + \smm\rho\spm \right) + \left(1 + \bar{n}(\omega,T)\right)\left( - \rho\smm\spm + \spm\rho\smm \right) \right\} \nn\\
&&+ e^{-i(\omega - \omega_{m+1,m}) t'} \left\{ \left(1 + \bar{n}(\omega,T)\right) \left(- \spm\smm\rho + \smm\rho\spm \right) + \bar{n}(\omega,T) \left( - \rho\smm\spm + \spm\rho\smm \right) \right\} \nn\\
&&+ e^{-i(\omega + \omega_{m+1,m}) t'} \left\{ \left(1 + \bar{n}(\omega,T)\right) \left(- \smm\spm\rho + \spm\rho\smm \right) + \bar{n}(\omega,T) \left( - \rho\spm\smm + \smm\rho\spm \right) \right\} \bigg].
\eea
Then, making use of the identity
\be
\int_0^\infty \id t e^{-i\omega t} = \pi \delta(\omega) - i \mathcal{P}\left(\frac{1}{\omega}\right), \label{EqDeltaPrincipal}
\ee
where $\mathcal{P}$ denotes principal value, we get after some work
\bea
\dot{\tilde{\rho}}(t) &=& 2\sum_{m} g_m^2 \bigg[ \pi J(\omega_{m+1,m}) \abssq{A(\omega_{m+1,m})} \bigg\{ \left(-\rho\spm\smm + \smm\rho\spm - \spm\smm\rho +\smm\rho\spm \right) \nn\\
&&+ \bar{n}(\omega_{m+1,m},T) \left( -\smm\spm\rho + \spm\rho\smm -\rho\spm\smm + \smm\rho\spm - \spm\smm\rho +\smm\rho\spm - \rho\smm\spm + \spm\rho\smm \right) \bigg\} \nn\\
&&+ i \mathcal{P}\int_0^\infty \frac{J(\omega) \abssq{A(\omega)}}{\omega - \omega_{m+1,m}} \bigg\{ \left(-\rho\spm\smm + \smm\rho\spm + \spm\smm\rho - \smm\rho\spm \right) \nn\\
&&+ \bar{n}(\omega,T) \left( -\smm\spm\rho + \spm\rho\smm -\rho\spm\smm + \smm\rho\spm + \spm\smm\rho - \smm\rho\spm + \rho\smm\spm - \spm\rho\smm \right) \bigg\} \nn\\
&&+ i \mathcal{P}\int_0^\infty \frac{J(\omega) \abssq{A(\omega)}}{\omega + \omega_{m+1,m}} \bigg\{ \left(-\rho\smm\spm + \spm\rho\smm + \smm\spm\rho - \spm\rho\smm \right) \nn\\
&&+ \bar{n}(\omega,T) \left( -\spm\smm\rho + \smm\rho\spm -\rho\smm\spm + \spm\rho\smm + \smm\spm\rho - \spm\rho\smm + \rho\spm\smm - \smm\rho\spm \right) \bigg\} \bigg] \nn\\
&=& 2\sum_{m} g_m^2 \bigg[ 2\pi J(\omega_{m+1,m}) \abssq{A(\omega_{m+1,m})} \bigg\{ (1 + \bar{n}(\omega_{m+1,m},T)) \lind{\smm}\rho + \bar{n}(\omega_{m+1,m},T) \lind{\spm} \rho  \bigg\} \nn\\
&&+ i \mathcal{P}\int_0^\infty \frac{J(\omega) \abssq{A(\omega)}}{\omega - \omega_{m+1,m}} \bigg\{ (1 + \bar{n}(\omega,T)) \comm{\ketbra{m+1}{m+1}}{\rho} - \bar{n}(\omega,T)  \comm{\ketbra{m}{m}}{\rho} \bigg\} \nn\\
&&+ i \mathcal{P}\int_0^\infty \frac{J(\omega) \abssq{A(\omega)}}{\omega + \omega_{m+1,m}} \bigg\{ (1 + \bar{n}(\omega,T)) \comm{\ketbra{m}{m}}{\rho} - \bar{n}(\omega,T)  \comm{\ketbra{m+1}{m+1}}{\rho} \bigg\} \bigg],
\eea
\end{widetext}
where we have introduced the notation $\lind{X}\rho = X\rho X^\dag - \frac{1}{2} X^\dag X\rho - \frac{1}{2} \rho X^\dag X$. Transforming back from the interaction picture and collecting terms yields the result given in Eqs.~(\ref{EqMasterEq})-(\ref{EqDefDeltam}).

As noted in the main text, the Lamb shift in \eqref{EqDefDeltam} diverges linearly for a small atom when $J(\omega)$ is ohmic. Bethe showed in the original calculation of the Lamb shift \cite{BethePR1947} how this can be remedied. Introducing the notation $q = \sum_{m}g_{m}\left(\smm + \spm\right)$, which in the case of a transmon is related to the charge on the island, note that \eqref{EqDefDeltam} in the limit of negligible temperature can be written
\bea
\Delta_m &=& -2\mathcal{P} \int_0^\infty \id \omega J(\omega)\abssq{A(\omega)} \sum_n \frac{\abssq{\brakket{m}{q}{n}}}{\omega+\omega_{n,m}} \nn\\
&=& -2\mathcal{P} \int_0^\infty \id \omega J(\omega)\abssq{A(\omega)} \nn\\
&&\times \left(\frac{m+1}{\omega + \omega_{m+1,m}} + \frac{m}{\omega - \omega_{m,m-1}} \right).
\eea
However, the renormalized electrostatic energy contribution from the atom, given by $q^2$, should already be incorporated in $\omega_m$. Thus we need to subtract
\bea
\Delta'_m &=& -2\mathcal{P} \int_0^\infty \id \omega J(\omega)\abssq{A(\omega)} \frac{\brakket{m}{q^2}{m}}{\omega} \nn\\
&=& -2\mathcal{P} \int_0^\infty \id \omega J(\omega)\abssq{A(\omega)}  \frac{2m+1}{\omega}
\eea
from $\Delta_m$. The result is the renormalized Lamb shift given in \eqref{EqDefDeltamRenorm}.


\section{Details of the (S,L,H) calculations}
\label{AppSLHCalc}

\subsection{Overview of (S,L,H)}

In this appendix, we do calculations in the $(S,L,H)$ formalism for cascaded quantum systems \cite{GoughCommMathPhys2009,GoughIEEE2009}. We first give a brief overview of the rules used in this formalism, following the supplementary material in \cite{KerckhoffLehnertPRL2012}. Each $(S,L,H)$ triplet represents a quantum system with a scattering matrix $S$, coupling operators forming a vector $L$, and a Hamiltonian $H$. There is a concatenation product $\boxplus$ (stacking channels) and a series product $\triangleleft$ (feeding output from one system into another): 
\begin{eqnarray}
G_2 \triangleleft G_1 &=& \bigg(S_2S_1, S_2L_1 + L_2, \nn\\
&&\quad H_1 + H_2 + \frac{1}{2i}\left(L_2^\dag S_2 L_1 - L_1^\dag S_2^\dag L_2 \right) \bigg), \quad\:\:\:\label{EqSeriesProduct}\\
G_2 \boxplus G_1 &=& \left( \begin{pmatrix} S_2 & 0 \\ 0 & S_1 \end{pmatrix}, \begin{pmatrix} L_2 \\  L_1 \end{pmatrix}, H_2 + H_1 \right). \label{EqConcatenationProduct}
\end{eqnarray}

There is also a rule for the feedback operation $[(S,L,H)]_{k\rightarrow l} = (\tilde{S},\tilde{L},\tilde{H})$, which represents feeding the $k^{th}$ output of a system into the $l^{th}$ input of the same system. The result is
\bea
\tilde{S} &=& S_{\slashed{[k,l]}} + 
\begin{pmatrix}
S_{1,l} \\
\vdots \\
S_{k-1,l} \\
S_{k+1,l} \\
\vdots \\
S_{n,l}
\end{pmatrix}
\left(1-S_{k,l}\right)^{-1} \nn\\
&& \times
\begin{pmatrix}
S_{k,1}\:
\dots \:
S_{k,l-1} \:
S_{k,l+1} \:
\dots \:
S_{k,n}
\end{pmatrix}, \\
\tilde{L} &=& L_{\slashed{[k]}} + 
\begin{pmatrix}
S_{1,l} \\
\vdots \\
S_{k-1,l} \\
S_{k+1,l} \\
\vdots \\
S_{n,l}
\end{pmatrix}
\left(1-S_{k,l}\right)^{-1}L_k,
\\
\tilde{H} &=& H + \frac{1}{2i}\left(\left(\sum_{j=1}^n L_j^\dag S_{j,l} \right)\left(1-S_{k,l}\right)^{-1}L_k - \text{h.c.} \right), \nn\\
\:
\label{EqFeedbackRule}
\eea
where $S_{\slashed{[k,l]}}$ and $L_{\slashed{[k]}}$ are the original scattering matrix and coupling vector with row $k$ and column $l$ removed.

Once we have the $(S,L,H)$ triplet for our total system, 
\begin{equation}
G = \left(S, \begin{pmatrix} L_1 \\ \vdots \\ L_n \end{pmatrix}, H\right),
\end{equation}
we can extract the master equation for the total system as
\begin{equation}
\dot{\rho} = -i\comm{H}{\rho} + \sum_{i=1}^n \lind{L_i}\rho,
\label{EqGeneralME}
\end{equation}
The output from port $i$ of the system is simply given by $\expec{L_i}$.

\subsection{Giant atom}

We begin by assigning a triplet for each connection point and each propagation direction. The coupling strength at connection point $k$ is denoted $\gamma_k$ and the phase shift between connection points $k$ and $k+1$ is $\phi_k=\omega_{1,0}(x_{k+1}-x_k)/v$. We only consider the case of a two-level atom to begin with. The phase shifts are accounted for by feeding the output from one connection point through a triplet $G_{\phi} = (e^{i\phi_k},0,0)$ before using it as input at the next connection point.

We will first look at the right- and left-travelling waves separately, and then combine the results. The triplet for the right-travelling wave at connection point $k$ is
\bea
G_{Rk} = \left(1,\sqrt{\gamma_k/2}\sm,0 \right),
\eea
except for $k=1$, where we also add the Hamiltonian $\frac{\Delta}{2}\sz$. We are working in a rotating frame where $\Delta = \omega_{1,0} - \omega_p$ is the detuning of the atom from some probe frequency $\omega_p$ we are interested in. Now, for $N=2$ the total triplet for the right-travelling waves can be written
\bea
G_{R,tot,2} &=& \left[\left(G_{\phi_1} \triangleleft G_{R1} \right) \boxplus G_{R2} \right]_{1\rightarrow 2} \nn\\
&=& 
\bigg( 
\begin{pmatrix}
e^{i\phi_1} & 0 \\
0 & 1
\end{pmatrix}
,
\begin{pmatrix} 
e^{i\phi_1} \sqrt{\gamma_1/2}\sm \\ 
\sqrt{\gamma_2/2}\sm
\end{pmatrix}
,
\frac{\Delta}{2} \sz
\bigg)_{1\rightarrow 2} \nn\\
&=& 
\left(e^{i\phi_1}, \bigg(\sqrt{\gamma_2/2} + e^{i\phi_1} \sqrt{\gamma_1/2} \right)\sm, \nn\\
&&\quad \frac{1}{2}\sz \left(\Delta + \frac{1}{2}\sqrt{\gamma_1\gamma_2} \sin(\phi_1) \right) \bigg).
\eea

Iterating this process gives the triplet for $N=3$,
\bea
G_{R,tot,3} &=& \left[\left(G_{\phi_2} \triangleleft G_{R,tot,2} \right) \boxplus G_{R3} \right]_{1\rightarrow 2} \nn\\
&=& \bigg(
e^{i(\phi_1+\phi_2)}, \nn\\
&&\left(\sqrt{\gamma_3/2} + e^{i\phi_2} \sqrt{\gamma_2/2} + e^{i(\phi_1+\phi_2)}\sqrt{\gamma_1/2} \right)\sm, \nn\\
&&\quad \frac{1}{2}\sz \bigg[\Delta + \frac{1}{2} ( \sqrt{\gamma_2\gamma_1} \sin(\phi_1) + \sqrt{\gamma_3\gamma_2}\sin(\phi_2) \nn\\
&& \quad + \sqrt{\gamma_3\gamma_1}\sin(\phi_2 + \phi_1) ) \bigg]
\bigg),
\eea
and by induction we arrive at the triplet for general $N$,
\bea
G_{R,tot,N} &=& \bigg(e^{i\phi_\Sigma}, A_R(\left\{\gamma_k,\phi_k\right\}) \sigma_-, \nn\\
&&\quad \frac{\Delta + \frac{1}{2}B(\left\{\gamma_j,\phi_j\right\})}{2} \sigma_z \bigg),
\eea
where we have defined
\bea
\phi_\Sigma &=& \sum_{k=1}^{N-1} \phi_k, \\
A_R(\left\{\gamma_k,\phi_k\right\}) &=& \sqrt{\gamma_N/2} + e^{i\phi_{N-1}}\sqrt{\gamma_{N-1}/2} \nn\\
&&+ e^{i(\phi_{N-1} + \phi_{N-2})}\sqrt{\gamma_{N-2}/2} + \ldots \nn\\
&&+ e^{i(\phi_{N-1} + \ldots + \phi_1)}  \sqrt{\gamma_1/2} \nn\\
&=& \sum_{j=1}^N \sqrt{\gamma_j/2} \exp\left(i\sum_{k=j}^{N-1} \phi_j \right), \\
B(\left\{\gamma_k,\phi_k\right\}) &=& \sum_{j=1}^{N-1} \sqrt{\gamma_j \gamma_{j+1}} \sin(\phi_j) \nn\\
&&+ \sum_{j=1}^{N-2} \sqrt{\gamma_j \gamma_{j+2}} \sin(\phi_j + \phi_{j+1}) + \ldots \:\:\:\:\:\: \nn\\
&&+ \sum_{j=1}^{2} \sqrt{\gamma_j \gamma_{j+N-1}} \sin \left(\sum_{k=j}^{j+N-2} \phi_k\right) \nn\\
&=& \sum_{i=1}^{N-1}\sum_{j=1}^{N-i} \sqrt{\gamma_j \gamma_{j+i}} \sin\left(\sum_{k=j}^{j+i-1} \phi_k\right).
\eea

We now turn to the left-travelling waves. The triplet for the left-travelling wave at connection point $k$ is
\bea
G_{Lk} = \left(1,\sqrt{\gamma_k/2}\sm,0 \right).
\eea
Now, for $N=2$ the total triplet for the left-travelling waves can be written
\bea
G_{L,tot,2} &=& \left[\left(G_{\phi_1} \triangleleft G_{L2} \right) \boxplus G_{L1} \right]_{1\rightarrow 2} \nn\\
&=& 
\bigg( 
\begin{pmatrix}
e^{i\phi_1} & 0 \\
0 & 1
\end{pmatrix}
,
\begin{pmatrix} 
e^{i\phi_1} \sqrt{\gamma_2/2}\sm \\ 
\sqrt{\gamma_1/2}\sm
\end{pmatrix}
,
0
\bigg)_{1\rightarrow 2} \nn\\
&=&
\bigg(e^{i\phi_1}, \left(\sqrt{\gamma_1/2} + e^{i\phi_1} \sqrt{\gamma_2/2} \right)\sm, \nn\\
&&\quad \frac{1}{2}\sz \left(\frac{1}{2}\sqrt{\gamma_1\gamma_2} \sin(\phi_1) \right) \bigg).
\eea
Carrying through the same procedure as for the right-travelling waves, we arrive at
\bea
G_{L,tot,N} &=& \bigg( e^{i\phi_\Sigma}, A_L(\left\{\gamma_k,\phi_k\right\}) \sigma_-, \nn\\
&&\quad \frac{\Delta + \frac{1}{2}B(\left\{\gamma_j,\phi_j\right\})}{2} \sigma_z \bigg),
\eea
where
\bea
A_L(\left\{\gamma_k,\phi_k\right\}) &=& \sqrt{\gamma_1/2} + e^{i\phi_1} \sqrt{\gamma_2/2} \nn\\
&&+ e^{i(\phi_1 + \phi_2)} \sqrt{\gamma_3/2} + \ldots \nn\\
&&+ e^{i(\phi_1 + \ldots + \phi_{N-1})} \sqrt{\gamma_N/2} \nn\\
&=& \sum_{j=1}^N \sqrt{\gamma_j/2}\exp\left(i\sum_{k=1}^{j-1} \phi_k \right).
\eea

Adding the left- and right-travelling waves, we thus have for general $N$ the total triplet
\bea
G_{tot,N} &=& G_{R,tot,N} \boxplus G_{L,tot,N} \nn\\
&=&
\bigg( 
\begin{pmatrix}
e^{i\phi_\Sigma} & 0 \\
0 & e^{i\phi_\Sigma}
\end{pmatrix}
,
\begin{pmatrix} 
A_R(\left\{\gamma_j,\phi_j\right\}) \sigma_- \\ 
A_L(\left\{\gamma_j,\phi_j\right\}) \sigma_-
\end{pmatrix}
, \nn\\
&&\quad \frac{\Delta + B(\left\{\gamma_j,\phi_j\right\})}{2} \sigma_z
\bigg),
\label{EqGeneralTripletDifferentCouplingsAndDifferentPhases}
\eea
We note that 
\bea
A_L &=& \left(A_R e^{-i\phi_\Sigma} \right)^*, \\
A_R &=& \left(A_L e^{-i\phi_\Sigma} \right)^*,
\eea
which entails $\abssq{A_L} = \abssq{A_R}$, and thus the relaxation rate, given in \eqref{EqRelaxSLH}, has the same frequency dependence as we saw from the derivation in \appref{AppContinuumCalc}. We also note that we here got a more explicit formula for the Lamb shift, $B(\left\{\gamma_j,\phi_j\right\})$, but it is equivalent to the result in \appref{AppContinuumCalc} with the added assumptions of negligible temperature, constant density of states $J(\omega) = J(\omega_{1,0})$, RWA on the level of the Hamiltonian, and extension of the lower integration limit in the $\omega$ integral to $-\infty$.

To extend the calculations to a multilevel giant atom, we need to add new channels for the higher transitions. However, since we assume large enough anharmonicity to avoid cross-talk between transitions, we can basically just reuse the result we just derived, taking into account the fact that the coupling increases with a factor $g_m$ for higher transitions. The result is still in agreement with that of \appref{AppContinuumCalc}.

\subsection{Giant atom in front of a mirror}

An interesting setup which is easily handled in the $(S,L,H)$ formalism is that of a giant atom placed in front of a mirror. Assuming that the mirror is close enough to the atom for travel times to be negligible, we can use the triplet from  \eqref{EqGeneralTripletDifferentCouplingsAndDifferentPhases} and modify it to our current situation by feeding the output from port 1 through a phase shift $\phi_M$ (representing the phase accumulated travelling to the mirror and back) and then feeding it back through port 2:
\bea
G_\text{mirror} &=& \left[ \left(G_{\phi_M} \boxplus I_1 \right) \triangleleft G \right]_{1\rightarrow 2} \nn\\
&=& 
\bigg( 
\begin{pmatrix}
e^{i(\phi_\Sigma + \phi_M)} & 0 \\
0 & e^{i\phi_\Sigma}
\end{pmatrix}
,\nn\\
&&\quad
\begin{pmatrix} 
e^{i\phi_M}A_R(\left\{\gamma_j, \phi_j\right\}) \sm \\ 
A_L(\left\{\gamma_j,\phi_j\right\}) \sm
\end{pmatrix}
,\nn\\
&&\quad
\frac{\Delta + B(\left\{\gamma_j,\phi_j\right\})}{2} \sz
\bigg)_{1\rightarrow 2} \nn\\
&=& \bigg( e^{i (2\phi_\Sigma + \phi_M)}, \nn\\
&&\left[ A_L(\left\{\gamma_j,\phi_j\right\}) + e^{i(\phi_\Sigma + \phi_M)}A_R(\left\{\gamma_j, \phi_j\right\}) \right] \sm, \quad\nn\\
&&\frac{1}{2}\sz\left(\Delta + B(\left\{\gamma_j,\phi_j\right\}) + \im\left(A_R^2 e^{i\phi_M}\right)\right)\bigg).
\eea
We thus have a modified relaxation rate 
\be
\Gamma_{1,0} = \abssq{A_L(\left\{\gamma_j,\phi_j\right\}) + e^{i(\phi_\Sigma + \phi_M)}A_R(\left\{\gamma_j, \phi_j\right\})}
\ee
and an addition of $\im\left(A_R^2 e^{i\phi_M}\right)$ to the Lamb shift, both depending on the relation between the distance to the mirror and the transition frequency (giving the phase shift $\phi_M$).

\end{document}